# Average power scaling of THz spintronic emitters in reflection geometry

**TIM VOGEL,**[1,*] **ALAN OMAR,**[1] **SAMIRA MANSOURZADEH,**[1] **FRANK WULF,**[1] **NATALIA MARTÍN SABANÉS,**[2,3] **MELANIE MÜLLER,**[4] **TOM S. SEIFERT,**[2] **ALEXANDER WEIGEL,**[5,6,7] **GERHARD JAKOB,**[8] **MATHIAS KLÄUI,**[8] **IOACHIM PUPEZA,**[5,6] **TOBIAS KAMPFRATH,**[2,4] **AND CLARA J. SARACENO**[1]

[1]*Photonics and Ultrafast Laser Science, Ruhr-University Bochum, 44801 Bochum, Germany*
[2]*Department of Physics, Free University of Berlin, 14195 Berlin, Germany*
[3]*IMDEA Nanoscience, Ciudad Universitaria de Cantoblanco, Faraday 9, 28049 Madrid, Spain*
[4]*Department of Physical Chemistry, Fritz Haber Institute of the Max Planck Society, 14195 Berlin, Germany*
[5]*Department of Physics, Ludwig-Maximilians-University Munich, Am Coulombwall 1, 85748 Garching, Germany*
[6]*Max-Planck-Institute of Quantum Optics, Hans-Kopfermann-Str. 1, 85748 Garching, Germany*
[7]*Center for Molecular Fingerprinting, Research Nonprofit LLC., Czuczor utca 2-10, 1093 Budapest, Hungary*
[8]*Institute of Physics, Johannes Gutenberg-University Mainz, 55128 Mainz, Germany*
*\*Tim.Vogel-u81@Ruhr-Uni-Bochum.de*

**Abstract:** Metallic spintronic THz emitters have become well-established for offering ultra-broadband, gapless THz emission in a variety of excitation regimes, in combination with reliable fabrication and excellent scalability. However, so far, their potential for high-average-power excitation to reach strong THz fields at high repetition rates has not been thoroughly investigated. In this article, we explore the power scaling behavior of tri-layer spintronic emitters using an Yb-fiber excitation source, delivering an average power of 18.5 W at 400 kHz repetition rate, temporally compressed to a pulse duration of 27 fs. We confirm that the reflection geometry with back-side cooling is ideally suited for these emitters in the high-average-power excitation regime. In order to understand limiting mechanisms, we disentangle the effects on THz power generation by average power and pulse energy, by varying the repetition rate of the laser. Our results show that the conversion efficiency remains mostly dependent on the incident fluence in this high-average-power, high-repetition-rate excitation regime if the emitters are efficiently cooled. Using these findings, we optimize the conversion efficiency to reach $5 \times 10^{-6}$ at highest excitation powers in the back-cooled reflection geometry. Our findings provide guidelines for scaling the power of THz radiation emitted by spintronic emitters to the mW-level by using state-of-the-art femtosecond sources with multi-hundred-Watt average power to reach ultra-broadband, strong-field THz sources with high repetition rate.

## 1. Introduction

Terahertz Time Domain Spectroscopy (THz-TDS) has become a ubiquitous tool in many fields of science and is being readily deployed in industrial settings [1–4]. While these systems are becoming more and more mature, cost-effective THz generation methods combining gapless broad bandwidth and high dynamic range (e.g., as provided by high THz average power and high repetition rate) remain rare.

Most industrial THz-TDS systems make use of semiconductor-based photoconductive emitters and receivers. These systems offer record-high dynamic range operation [5], and the corresponding emitters provide high conversion efficiencies with low power excitation.

Typically, these systems are driven by robust fiber lasers operating at repetition rates in the hundreds of MHz regime. However, they are limited mainly by carrier lifetimes to bandwidths <6 THz, and high pulse energy excitation for high-THz-field applications is typically limited by two-photon absorption, with the exception of few specially designed large-area emitters [6]. State-of-the-art results recently showed the generation of THz average power of 637 µW at 100 MHz with a bandwidth of 6 THz [5]. Using plasmonic structures even milliwatt-class THz average power was reached using 720 mW excitation with 4 THz bandwidth [7]. Higher average-power excitation schemes remain to be demonstrated using specialized large-area emitters, which are, however, not straightforward to engineer.

Optical rectification in a nonlinear crystal with broken inversion symmetry, for instance zinc telluride (ZnTe), gallium phosphide (GaP), lithium niobate ($LiNbO_3$) or organic crystals (e.g., BNA, DAST, DSTMS) is a common technique for THz-TDS. This technique is typically well suited for higher pulse-energy excitation and recently demonstrated its potential for average power excitation in the multi-100-W average-power regime, yielding average THz powers well beyond the milliwatt level [8–11]. However, phase-matching conditions and phonon resonances limit the generation of wide, spectrally flat bandwidths and the application of high energy levels is also often limited by multi-photon absorption.

Two-color plasma filaments allow to generate ultra-wide bandwidths, where high conversion efficiencies in the percent level have been reached free of material damage [12]. This technique is potentially interesting for high-average power excitation, with recent results achieving 640 mW THz average power at 500 kHz and spectral content up to 30 THz [13]. Applications such as molecular spectroscopy benefit from broadband THz sources, e.g. to capture vibrational modes of organic molecules [14]. However, expensive and complex laser amplifier systems are typically needed to reach millijoule-level energies and sufficiently high peak intensities required to achieve a highly ionized plasma [15]. Therefore, such sources remain mostly restricted to laboratory applications. Furthermore, it appears that at much higher repetition rates, a compromise needs to be found in terms of conversion efficiency due to accumulation effects [16]. Moreover, the combination of broad bandwidths in these setups with high dynamic ranges remains to be demonstrated, even with a high repetition rate.

Spintronic THz emitters (STE) combine many of the advantages of the above-mentioned methods [17,18]. They are based on optically generated spin currents and subsequent spin-to-charge-current conversion in nanometer-thin magnetic metal layers. This approach results in single-cycle THz pulses with a gapless flat spectrum from 1 THz up to 30 THz, limited by the input pulse duration for pulses longer than ~ 10-15 fs [19]. In contrast to plasma sources, the THz field from STE scales linearly with the pump fluence in the low-excitation regime. Thus, low-energy pulses at high repetition rates as well as high-energy ultrafast lasers can be used for excitation as long as the fluence remains below the damage threshold, opening the door to more widely-applicable broadband THz sources. Another benefit of STEs is their reliability, homogeneity and upscalability of the production, since depositing thin metal layers without any structure/lithography is a mature technology. The size of the STE can be upscaled to large apertures for ultra-high energies, and samples up to 7.5 cm have been already fabricated [20]. An exhaustive study of the influence of different metal layers was conducted and rather than the common bi-layer system, the tri-layer system offered signal levels similar to the ones obtained by standard THz semiconductor emitters, like GaP or ZnTe [19]. Because the STE consists of metallic films that are only a few nanometers thick, no phase-matching condition needs to be fulfilled, allowing one to operate the STE in transmission or reflection geometry simultaneously, in stark contrast to most other THz sources.

So far, STEs have mostly been excited by with ultrashort laser pulses from sources with limited average power based on Ti:Sapphire modelocked oscillators at ~80 MHz repetition rates and pulse energies in the nanojoule regime, or amplifiers with few kHz repetition rates and millijoule pulse energies. For example, up to 30 THz bandwidth have been demonstrated for a pump power of 100 mW and a repetition rate of 80 MHz [19], and up to 10 THz were demonstrated at 1 kHz repetition rate and 5 W of pump power [20]. Other pumping regimes were also demonstrated [21], mostly exploiting the independence of STEs in terms of excitation wavelength (400 nm [22], 800 nm [23], as well as 1000 nm to 1300 nm [24] and 1550 nm [25]). In this context, an important, yet largely unexplored, pumping regime of STEs is the high-average-power, high-repetition-rate regime as provided by modern Yb-based laser sources [26–29]. The thin metallic structure has inherently excellent thermal properties, and it can be expected that heat from the pump laser is efficiently dissipated by proper heat sinking.

In this article, we explore excitation of STE in the high-average-power, high-repetition-rate regime. We study different cooling geometries, including the promising reflection geometry, for the first time in the context of average power scaling. As a pump source, we used a commercial fiber-laser that delivers up to 18.5 W of average power, temporally compressed to sub-30 fs to explore the average power scaling of a tri-layer STE (TeraSpinTec GmbH) on a sapphire support substrate, which was shown to exhibit good thermal properties [30]. By varying the repetition rate from 400 kHz down to 40 kHz, we disentangle the effect of average power and pulse energy on the THz generation efficiency. We also study the impact of cooling in reflection geometry versus a free-standing configuration, which can be compared to the more common transmission geometry. We find that for higher pump powers, backside cooling gives a benefit in terms of generated average THz power, and allows to pump the emitter harder without degradation in efficiency.

Additionally, we show that the optimally cooled STE exhibits a performance independent of the repetition rate, as long as the pump beam diameter is adjusted for an optimal fluence. Our results allow us to establish first guidelines towards pumping these novel THz emitters with multi-100 W laser systems at even higher repetition rates in optimized cooling conditions, which are accessible, for example, with state-of-the-art thin-disk oscillators [31].

## 2. Experimental setup and methods

### 2.1 Laser system and pulse compressor

An overview of the experimental setup is given in Fig. 1. As a driving source, we use a commercial fiber laser (Trumpf TruMicro 2000) that delivers up to 18.5 W of average power at 1030 nm, with a pulse duration of 310 fs. The laser repetition rate can be varied at constant pulse energy between 400 kHz and 40 kHz, thus correspondingly changing the average power impinging on the STE sample. We have now two parameters - optical attenuation and change of the repetition rate - that allow us to study peak-power-dependence and thermal effects independently.

To exploit the potential of STEs for broad bandwidths, the shortest possible excitation pulses are required [19]. Therefore, we temporally compress the output pulses of the laser using an efficient Herriott-type multi-pass cell (MPC) compressor [32] to bring the pulse duration to the sub-30 fs regime, guiding this laser system to pulse durations typically available from Ti:Sapphire systems. MPC-based compressors enable spatially homogeneous spectral broadening of the pulses in a free-space setup while keeping the beam quality high, where the pulse can afterwards be compressed to a nearly Fourier-limited duration by dispersive mirrors. This scheme has shown outstanding performance for bringing long-pulse Yb-based setups into the sub-100 fs regime [33–35], enabling them to be used for applications typically reserved to Ti:Sapphire laser systems. In addition to high compressibility, MPC compressors preserve the spatial beam quality, with power scaling from few watts to kilowatts [36], and achieve high optical transmission compared to other external compression techniques.

The Herriott-type MPC used in this experiment consists of two high-reflectivity-coated plano-concave mirrors with a radius of curvature of 300 mm and a separation of 540 mm. This configuration provides 21 roundtrips, i.e., 42 passes through the nonlinear medium, a 6-mm-thick anti-reflective-coated fused silica (FS) plate, placed approximately 100 mm apart from one of the MPC mirrors. One of the two mirrors has a group-delay dispersion (GDD) of -350 fs$^2$ per bounce, which compensates for dispersion of the 6-mm fused silica plate, but also adds additional negative dispersion, resulting in total in a slightly negative dispersion in the cell. We numerically optimize the position and the thickness of the FS, as well as the number of passes to obtain the highest peak power at the output. Final compression is achieved by using a pair of dispersive mirrors, adding a total GDD of -1250 fs² compressing the pulse close to its Fourier limit.

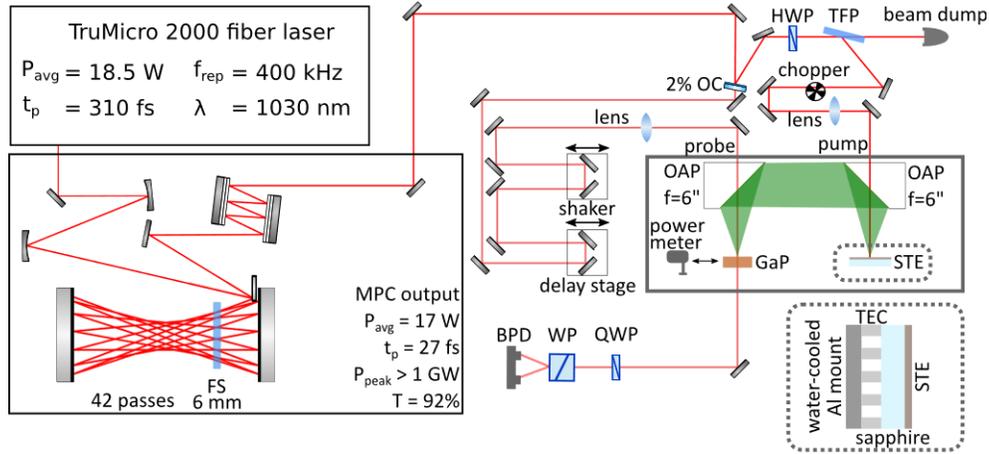

Fig. 1 THz-TDS setup in the case of the STE used in reflection geometry. The free-standing STE, as generally used in transmission geometry, was also studied and is schematized in Fig. 3 below. FS: Fused silica, MPC: Multi Pass Cell, T: Transmission, HWP: Half wave plate, TFP: Thin film polarizer, OC: Output coupler, OAP: Off-axis parabolic mirror, f: focal length, STE: Spintronic THz emitter, GaP: Gallium phosphide, QWP: Quarter wave plate, WP: Wollaston prism, BPD: Balanced Photodetector. Inset surrounded with dashed lines: Side-view of STE and mount. The STE consists of three thin metal layers on top of a sapphire substrate. With thermal paste on the backside of the substrate it is placed on the cold side of a thermoelectric cooler (TEC/Peltier element). The hot side is in contact with a water-cooled aluminum (Al) mount.

We characterized the compressed pulses using a second-harmonic-generation frequency-resolved optical gating (SHG-FROG) setup [37]. The results are presented in Fig. 2. The measured and reconstructed traces [Fig. 2a) and b)] are in good agreement with each other, with a calculated FROG error of 0.5% in a 1024×1024 grid. Fig. 2c) shows the measured spectrum after the compression stage using an optical spectrum analyzer together with the reconstructed spectrum and spectral phase from the FROG measurement. The excellent agreement between reconstructed and independently measured spectra shows the fidelity of the retrieved pulses. The pulse intensity profile [Fig. 2d)] is close to the Fourier-transform limit. The compressed pulse has a peak power of 1 GW and an intensity full width at half maximum (FWHM) pulse duration of 27 fs, which were calculated from the intensity profile obtained from the FROG measurement. The corresponding temporal phase of the pulse is plotted (grey dashed line), showing some residual phase remaining due to the limited bandwidth of the only available dispersive mirrors at the time of the experiments.

The transmission of the MPC including the compression mirrors is 92% and results in a compressed average power of 17 W available for the experiment. Throughout the THz generation experiment, we use a chopper wheel with a duty cycle of 50% in the pump arm as required for lock-in detection. It reduces the average power arriving at the emitter by half, which is also typically the case in previous experiments using Ti:Sapphire lasers [38]. Additionally, the laser system has an integrated Pockels cell making it possible to divide the repetition rate by an integer but keeping the pulse energy constant. The highest repetition rate available is 400 kHz, but in our investigation we vary the repetition rate down to 40 kHz, which allows us to reduce the average pump power while keeping the pulse energy and pulse duration constant, making it possible to disentangle the influence of average power and pump energy on the THz generation.

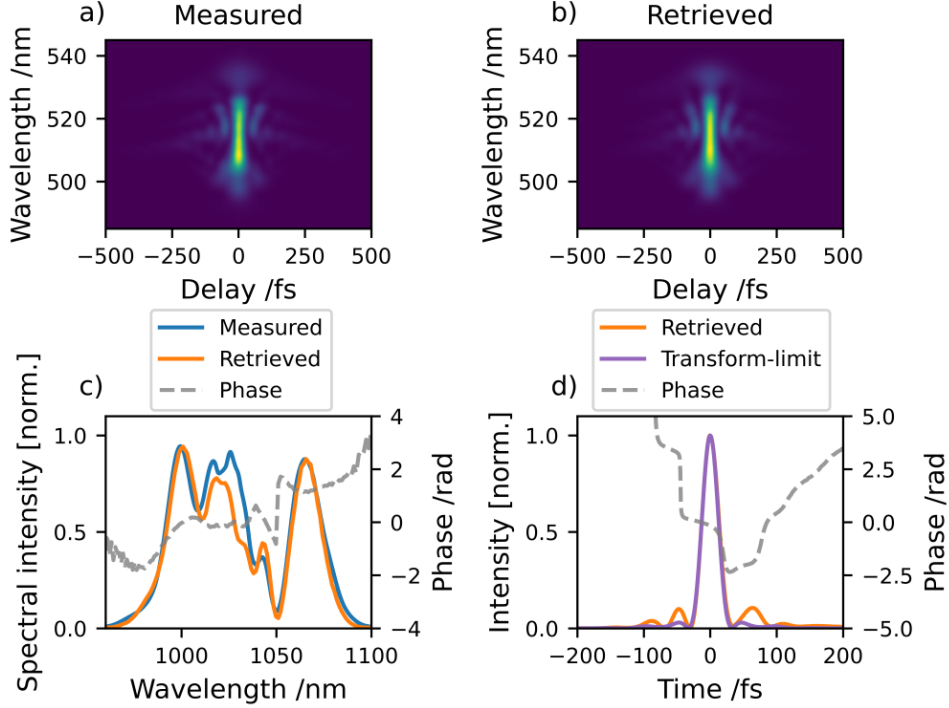

Fig. 2 Characterization of laser pulse after MPC. a) Measured FROG trace. b) Retrieved FROG trace on a grid of 1024 × 1024 and an error of < 0.5%. c) Measured spectral intensity with optical spectrum analyzer as well as the retrieved spectrum from FROG. The phase is displayed as dashed, grey line d) the retrieved pulse duration is close to the transform limit (green) of the given bandwidth.

*2.2 THz detection setup and STE*

The THz detection setup is presented in Fig. 1 for the case of the STE used in reflection geometry. For detection of the THz electric field in the time domain via electro-optic sampling (EOS), a 2% output coupler (OC) splits the beam into pump and probe. The high-power fraction of the beam is directed towards a half-wave plate in a computer-controlled motorized rotation mount in connection with a broadband Thin Film Polarizer (TFP), which makes it possible to control the average power (i.e. the pulse energy at a fixed repetition rate) on the STE in a precise and reproducible way. A chopper wheel is used to modulate the pump beam with either 18 Hz (necessary for the THz power meter) or 1337 Hz (modulation frequency for the EOS).

The STE used throughout this study (bottom-right inset Fig. 1) is a 5.8 nm thick metallic tri-layer structure W|$Co_{40}Fe_{40}B_{20}$|Pt, deposited on a 500 µm thick sapphire substrate with 75 mm diameter. It is mounted with thermal paste onto a Peltier element, and, for further rigidity, two-component epoxy is used at the sides. In this way, we can efficiently cool the thin structure from the backside. In order to evaluate the effect of active backside cooling on performance, we also tested the reflection geometry with active cooling off and the more common transmission geometry, where the laser can pass through the emitter. In the first two cases, the backside is contacted with thermal paste to a water-cooled Peltier element with

adjustable temperature. The emitter can be moved along three orthogonal axes with linear stages and is placed at the focus of the first off-axis parabolic mirror (OAP). A bar magnet (Nickel, N40) is placed horizontally on the table and under the STE to saturate the in-plane magnetization. The external magnetic field is measured with a gauss meter (Model 750D, RFL Industries, Inc.) and has a strength of 48 mT at the position of the pump spot, which is more than the threshold of 10 mT to saturate the magnetization of the STE [19].

The pump laser traverses a 3 mm aperture (parallel to the focused beam) of a 76.2 mm diameter OAP with $f$=152.4 mm focal length. The beam diameter on the STE is either set by a single plano-convex lens or a telescope consisting of two lenses. The diameter is measured with a CMOS camera in the focus of the OAP. The emitted THz radiation is then collimated with the aforementioned OAP and focused by a second identical OAP. The distance between the two OAPs is $2f$, resulting in a $4f$ 1:1 imaging telescope.

The EOS traces are measured using 200 µm-thick GaP as the detection crystal. We note that we focus our attention here on understanding the scaling of the average power of the STE and thus not necessarily on measuring the widest bandwidths in this experiment. The detection bandwidth could be improved in our setup by using thinner GaP and/or GaSe crystals [39,40]. The probe beam is guided over a delay stage with a mounted retroreflector to adjust the temporal overlap between pump and probe. A fast delay stage (ScanDelay 15 ps, APE GmbH) is used for recording traces. Similar to the pump beam, the probe beam is guided through a hole in the second OAP and focused on the detection crystal. The beam is routed through a quarter waveplate and a Wollaston prism. A balanced photo detector receives the two beams.

To measure the power emitted by the STE, a calibrated THz power meter (THz20, SLT GmbH) is used. The sensor was calibrated at the "Physikalisch-Technische Bundesanstalt" (PTB, German metrology institute) for THz power at 1.4 THz. According to the specification of the manufacturer, the sensitivity of the THz power meter fluctuates less than $\pm 5\%$ from 0.6 THz to 6 THz. In order to ensure that the measured power results only from the THz radiation and not residual pump light, either a high-density polyethylene (HDPE) filter or a thin sheet of Polytetrafluoroethylene (PTFE) is placed in the collimated beam. The filter calibration, measured with the THz power meter, shows a transmission of 80% for HDPE and 93% for PTFE. The power meter itself has a black textile over its input aperture with a transmission of 71%. Additionally, the setup is housed in a box and covered with plexiglass that can be purged with nitrogen to reduce absorption of THz by water. The dewpoint in the setup is at $(9 \pm 2)°C$, measured with a humidity sensor (SF-72, Michell Instruments), and limits very strong cooling in our experiment due to possible condensation on the emitter.

We also measure the surface temperature with a thermal camera (VarioCam HD, InfraTec). Due to the semi-reflective surface of the STE, the exact emissivity is difficult to evaluate. However, since temperatures at low pump powers agree with the water temperature in the actively cooled case, we use an emissivity $\epsilon = 1$ throughout this investigation. This is a reasonable assumption since we are mostly interested in temperature changes, and deviations in the absolute temperature values are uncritical to the conclusions. A different calibration file of the thermal camera was needed for temperatures above 120°C. The chopper is modulating the pump beam with 18 Hz, and a lock-in amplifier is used to extract the measured power and suppress noise from other sources.

## 3. Results

The experimental section is structured as follows: First, we explore the influence of cooling the STE in reflection geometry on the generated THz power and temperature of the emitter by comparing it to an uncooled geometry typically used in transmission, and confirm that at highest average powers, cooling is efficient and scaling laws are power independent. Once we established the best cooling geometry, we explore saturation of the STE by varying (1) the pump-pulse energy (via the average power at a fixed repetition rate) and (2) the repetition rate at a constant pump-spot diameter of 1.1 mm (intensity $1/e^2$), allowing us to observe saturation of the emitter in our parameter range. These measurements show that the fluence on the emitter – and not the average power – is responsible for emitter saturation. Finally, we optimize the emitted THz power by fixing the repetition rate at the maximum value of 400 kHz and varying the fluence via the spot size to operate at highest possible conversion efficiency for maximum generated THz power. We find that for our laser with a maximum pulse energy of 35 µJ, a pump diameter of 1.8 mm ($1/e^2$) results in an efficiency peak at the highest pump power of 7 W (on the STE). We additionally present a simple model, which discusses these results and gives insight into the scaling law of the STE. In these optimal excitation conditions, we characterize the THz radiation of the STE (such as EOS, polarity) and explore its thermal handling in more detail.

### 3.1 Effect of STE cooling

The different schematics displayed in Fig. 3g) – i) show the three different STE cooling geometries evaluated in this study. Fig. 3g), refers to a free-standing STE, a configuration which can be best compared with the commonly used transmission geometry. In this case, the STE is held from the side, such that the beam can propagate through the emitter and substrate. Fig. 3h) and Fig. 3i) show the cases where the STE is used with a support structure on the backside. In the case illustrated in Fig. 3h) we do not actively cool the emitter and in Fig. 3i) we actively cool the STE in reflection.

Fig. 3a) – c), shows the obtained THz power as a function of fluence for different repetition rates in these three cooling geometries. In the free-standing case [Fig. 3a)], saturation is observed beyond a certain fluence threshold for all repetition rates. The fluence at which this point occurs decreases with higher repetition rate indicating an effect related to the increased average power, that is, an accumulated heating. However, this saturation fluence does not appear to scale linearly with average power indicating other limiting mechanisms are in place, too. Beyond that point, the saturation behavior also depends on the repetition rate: in the case of highest repetition rate (400 kHz), a strong drop in the obtained power is visible, illustrating the onset of strong thermal effects which ultimately degrade the metallic structure [41]. For decreasing repetition rates, this degradation only starts at higher fluences (200 kHz) and ultimately results in a saturated, constant power (133 kHz). Fig. 3b) and c) show the cases where the STE is used in reflection with improved heat dissipation. In Fig. 3b) we do not actively cool the emitter, but the heat is still dissipated significantly better: once saturation is reached, the power is saturated but does not degrade even at highest repetition rates, in a similar way to the 133 kHz case in free-standing geometry. Specifically, THz power saturation occurs at nearly constant fluence, with no effect of the average pump power, indicating that THz emission efficiency is limited by the energy density within a single pulse. In Fig. 3c) the optimal cooling geometry is shown: the threshold for power saturation is shifted to higher fluences, so that higher THz powers can be reached. Nevertheless, we observe that the power saturation occurs in all configurations with only a slight shift to higher fluences when the structure is better cooled. At maximum input power (7 W on the STE), the measured Peltier cooling power is 4.2 W. A previous study showed that the tri-layer STE with a total metal thickness of 5.8 nm, like in our investigation, has an absorption of ~50%, a transmittance of ~35%, and a reflection of ~15% [19]. We observe with a thermopile power meter behind the STE a transmitted pump power of 33%, measured in the free-standing configuration and in agreement with literature. The cooling power of 4.2 W is therefore comparable to the non-reflected pump power of up to 6 W (circa 85% of the full incident average power).

In order to clarify if the observed power saturation is related to the absolute temperature of the structure, the maximum temperature on the STE is measured for all structures for various repetition rates in Fig. 3d) – f). A linear fit is applied to the temperature data and the base temperature (without pump power) for the free-standing case is $(18.9 \pm 3.2)$°C and for the deactivated thermo-electric cooling $(24.6 \pm 1.3)$°C. In the case of the active thermo-electric cooling, the base temperature is $(11.4 \pm 1.0)$°C. Contrarily to the power curves, the temperature increases linearly with the power impinging on the structure for all repetition rates as expected in a regime of linear absorption. The temperature slopes for the free-standing case in Fig. 3d) are $(16.3 \pm 1.2)$°C/W. With deactivated cooling from the Peltier element [Fig. 3e)], the temperature also rises linearly with increasing incident power, albeit with a smaller slope of $(9.8 \pm 0.5)$°C/W.

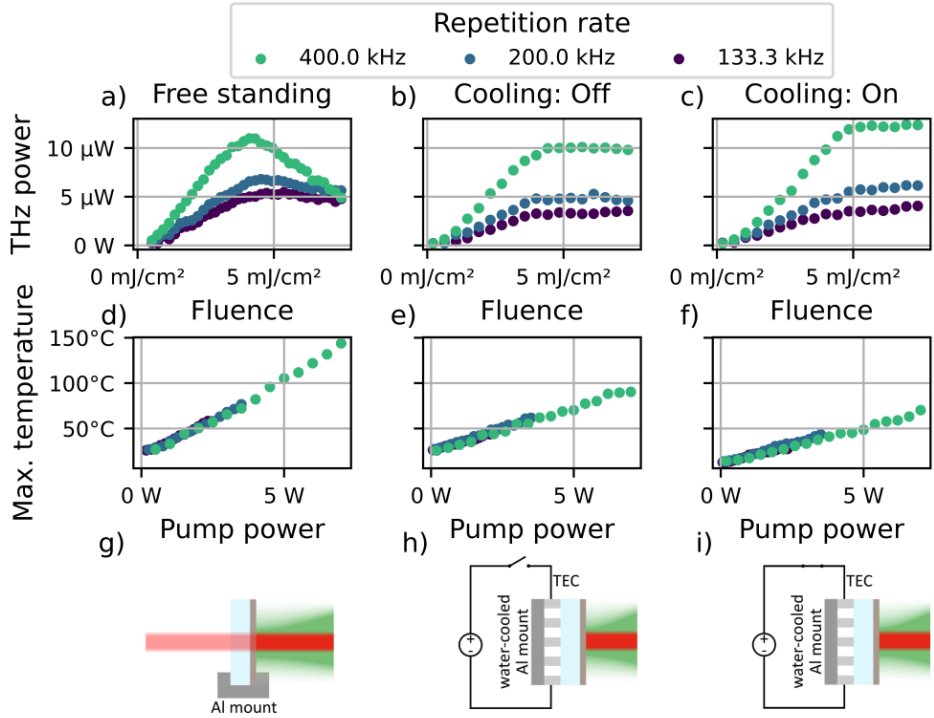

Fig. 3 a) – c) THz power versus fluence for various repetition rates. d) – f) Maximum temperature of the STE versus pump power for various repetition rates. g) Schematic of the free-standing configuration, the pump light can leave the substrate on the backside. This is close to the common transmission geometry which is typically used for THz generation. h) STE mounted on Peltier element/TEC (thermo-electric cooler), which is off, and an aluminum mount (without water flow). i) STE still mounted on Peltier element, this time with active cooling (approximately 4 W). The pump laser (red) propagates from the right and gets partly on the surface of the STE reflected. In all configurations is THz (green) collected opposite to the incoming laser beam. The pump diameter is in all configurations 1.1 mm ($1/e^2$).

The lowest increase in temperature for the applied pump power is observed for the actively cooled STE in Fig. 3f with $(8.1 \pm 0.6)$°C/W. Our measurements show on the one hand that the power saturation does not seem to be entirely related to reaching a critical temperature, since for a given cooling geometry at lower repetition rates, the saturation occurs at significantly lower absolute temperatures, i.e. the fluence at which this saturation occurs does not scale linearly with repetition rate. Nevertheless, lowering the temperature of the structure allows us to reach slightly higher fluences before saturating at the highest repetition rate of 400 kHz. This points to two mechanisms acting simultaneously: one related to the steady-state temperature of the emitter, i.e. scaling with average power, and another one scaling with fluence or peak power, such as transient heating of the metallic film [42]. Disentangling these two effects will be the subject of a separate investigation.

One practical consequence of our findings remains that active back-side cooling allows for smaller average temperature rises and to pump the emitters at higher average powers. Independently of the fact that the power saturation is observed with only a small dependence on average power, operation at lower absolute temperature ensures better stability and mitigates long-term degradation of the metallic structure. Furthermore, we show that the influence of cooling on achieving highest efficiency and output THz powers becomes more relevant for the highest repetition rates, which has important consequences for future experiments at higher average power and repetition rates in the MHz regime. In this case, better cooling will become critical, for example by using diamond heatsinks and lower cooling temperatures. With these measures we expect that high pumping average powers at high fluences without saturation can be reached.

*3.2 Exploration of STE saturation behavior*

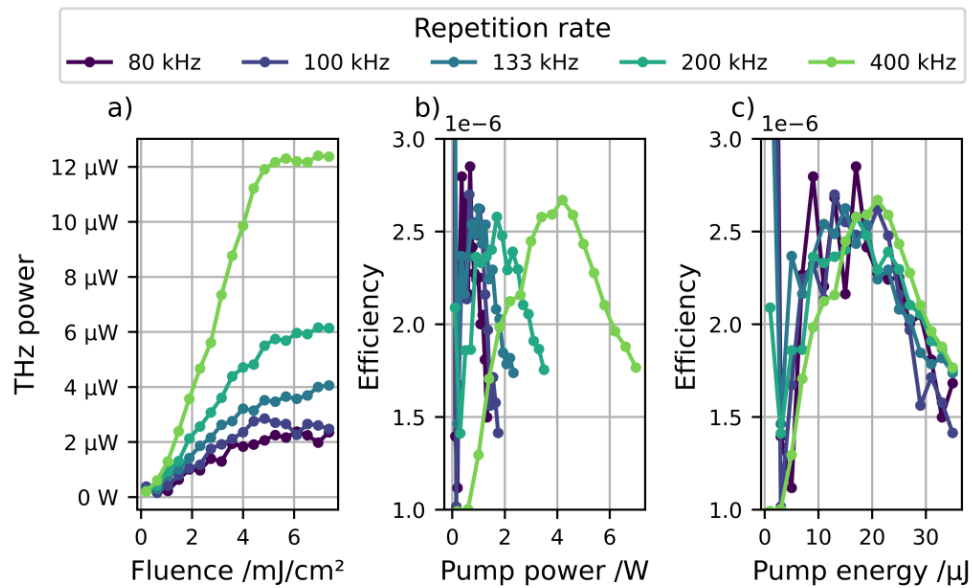

Fig. 4 a) THz power versus fluence for various repetition rates. b) Efficiency of THz generation versus pump power for various repetition rates. c) Efficiency versus pump energy for various repetition rates. The pump diameter is in all cases 1.1 mm ($1/e^2$).

In this section, we explore the observed power saturation from the previous section in more detail using the best configuration for cooling, i.e., the actively back-cooled reflection geometry. We vary pulse energy and repetition rate in a wider range at a constant pump-spot diameter of 1.1 mm ($1/e^2$), allowing us to saturate the STE. The results are presented in Fig. 4.

Fig. 4a) shows the absolute THz power versus fluence for different repetition rates of the laser [including the data presented in Fig. 3c)]. When changing from 400 kHz to 80 kHz, the maximum pulse energy remains unchanged, but the average pump power reduces, which in turn results in lower average THz power. The maximum THz power of 12 µW can be reached with the highest repetition rate, corresponding to the highest incident average power of 7 W (on the STE). In all cases, the THz power saturates at a constant fluence of 5 mJ/cm², thus limiting the THz power to lower values for low repetition rates. The maximum THz pulse energy is approximately 30 pJ and is reached at all tested repetition rates. The maximum peak intensity for these settings is 210 GW/cm². Increasing the fluence above 7 mJ/cm² results in permanent damage of the STE. Fig. 4b) allows us to study this saturation effect: we plot optical-to-THz conversion efficiency as a function of pump power for various repetition rates. For each repetition rate, an optimal average power operation point can be found corresponding to a fluence of 5 mJ/cm² mentioned above. With decreasing repetition rate, the efficiency peak moves to lower pump powers accordingly. Note that the slight increase in efficiency curves for low pump powers (<0.5 W) is assigned to the noise floor of the THz power meter (approximately 1 µW). The pump energy is unaffected by the repetition rate change from the Pockels cell, and each repetition rate is tested with up to 35 µJ of pulse energy. Fig. 4c) further illustrates the optimum operation point of the emitter by plotting the efficiency as a function of pulse energy (rather than power), showing approximately the same optimal pulse energy for all repetition rates. All tested repetition rates show a consistent efficiency peak of $2.5 \times 10^{-6}$ for 20 µJ of pump energy.

Our measurements show that in our excitation regime, the peak efficiency only depends on the fluence and is independent of the repetition rate, at least in the average power regime explored in this investigation and when cooling is efficient. As shown in the previous paragraph, this can be mostly attributed to the good thermal management in the back-cooling geometry. In our experiment, we could not easily vary the pump pulse duration within a wide range. In future experiments, this will also be an interesting parameter to vary, to further decouple fluence from intensity dependent effects. This will ideally complement future experiments where the transient dynamics of the observed effects could be studied.

### 3.3 Optimization of output average power

After finding that the maximum THz power is limited by a pulse fluence of about 5 mJ/cm², we optimize our experimental setup for output average power by varying the pump diameter from 0.65 mm up to 1.8 mm. For this subsection, the laser was kept at its highest repetition rate of 400 kHz, delivering the maximum average power available.

In Fig. 5a), the emitted THz power for various beam diameters is shown and in Fig. 5b) the corresponding efficiency is displayed. For a pump diameter of 0.65 mm ($1/e^2$), we observe that the measured THz power saturates at pump powers just above 1 W. The THz power does not increase above 3 µW, barely above the detection limit of the power meter. Increasing the beam diameter to 1.1 mm does not only show a higher saturation threshold of approximately 4 W, as in the previous subsections, but also results in higher THz power compared to the smaller spot size. Increasing the diameter further moves the peak in the efficiency curve to higher pump powers, as expected, and nearly optimum conversion is reached for a pump diameter of 1.8 mm at the maximum applied power of 7 W. The pump diameter adjustment with the telescope was restricted to the available lenses at the time of the experiment. When purging the setup with dry nitrogen to reduce the absorption of THz in the water vapor in the air, we observe a further increase of approximately +17% in average THz power. The apparent increase in efficiency for relatively low pump powers (<1 W) arises from the noise floor of the THz power meter (see above). At the maximum input power of 7 W, we obtain a THz average power of 39 µW, corresponding to a conversion efficiency of $5.6 \times 10^{-6}$.

The 50% duty cycle chopper wheel used in our setup allows for optimal noise suppression both for power measurements but also for EOS, and thus higher dynamic range in our setup. At the pulse repetition rate of 400 kHz (and lower), the temperature in the metal is expected to relax from pulse to pulse. In fact, whereas electronic temperatures in metal relax with ps timescales, electron phonon equilibrium is only reached is ns time-scales which is still much faster than the time between two pulses. Therefore, accumulation effects are unlikely to happen, and the conclusions made about thermal effects in the case of STEs can be applied to other duty cycles (including the potential case of no chopper wheel) [42], in this repetition rate range. However, this will be important to consider for future experiments at multi-MHz repetition rates, where accumulation effects will make the burst length a critical parameter. It is worth noting that this burst length independent operation is fundamentally different from the case of broadband nonlinear organic crystals used for optical rectification, where chopping is critical to let the crystal cool down between bursts, at correspondingly moderate repetition rates [11]. In the case of a thin metallic sheet, these accumulation effects will only start playing a role at much higher repetition rates. It was recently shown [43], that the manipulation of the magnetic field allows tailoring the THz polarization. Using an electro-magnet instead of a static magnetic field, the THz radiation can also be modulated [44]. That would enable lock-in detection also at full pump power without the chopper wheel and permit the use of the full repetition rate of the laser, making pump powers of up to 14 W possible.

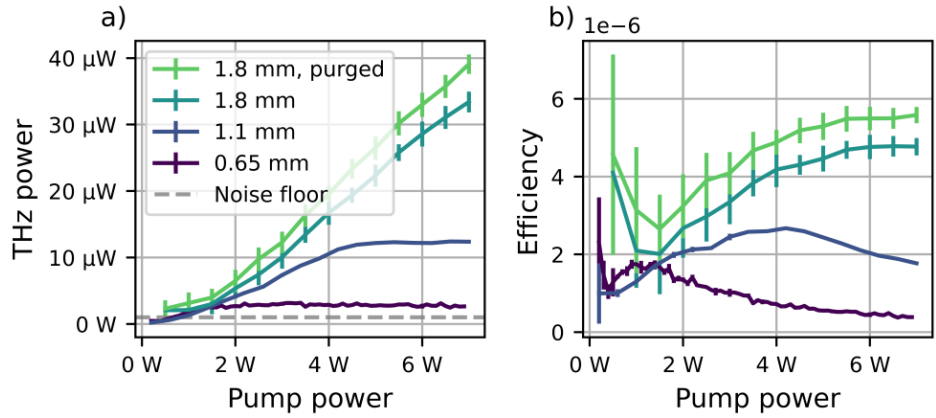

Fig. 5 a) THz power versus pump power for various pump diameters ($1/e^2$) at the maximum repetition rate of 400 kHz. For 1.8 mm diameter the setup is also purged with dry $N_2$, which results in a dewpoint of $(-28 \pm 2)$°C. The noise floor of the power meter in connection with a lock-in amplifier is approximately 1 µW. b) Corresponding efficiency of the STE. The beam diameter of 1.8 mm has the efficiency peak at the highest applied power of 7 W (on the STE). The higher efficiency for the lowest pump power is a measurement artifact due to the noise floor of the power meter.

After finding the optimum pump diameter of 1.8 mm ($1/e^2$ intensity) for a maximum pump energy of 35 µJ, we use this spot size in the following subsections.

### 3.4 Characterization of THz electric field

Fig. 6a) shows two time traces for opposite STE magnetizations and Fig. 6b) their corresponding power spectra recorded with EOS using a 200-µm-thick GaP crystal and a pump power of 7 W. Turning the external magnet by 180° flips the magnetic field and in turn reverses the polarity of the trace in the time domain. The results of this polarization dependent measurement agree with the expectation. Additionally, we find that the spectra of both traces are in good agreement, with the dips being absorption from water vapor in the air. No further processing or filtering is done to the frequency data.

The spectral bandwidth, when measured up to the frequency where the noise floor starts, extends up to 6 THz. In the measurement of Fig. 6b), the limited bandwidth is due to in big part to the limited phase matching bandwidth of the 200 µm thick GaP detection crystal at 1030 nm, low pass filtering of the lock-in amplifier in connection with a sinusoidal moving stage and the absorption peak of the PTFE filter at 6.1 THz. However, our attempts to measure a wider bandwidth of the emitted THz radiation using thinner GaP, GaSe or our FTIR did not significantly improve the bandwidth, most likely because of the reduced dynamic range in these cases. Since the goal of this study was to explore mostly power handling and efficiency, the current setup allowed us for maximum flexibility and ease of detection.

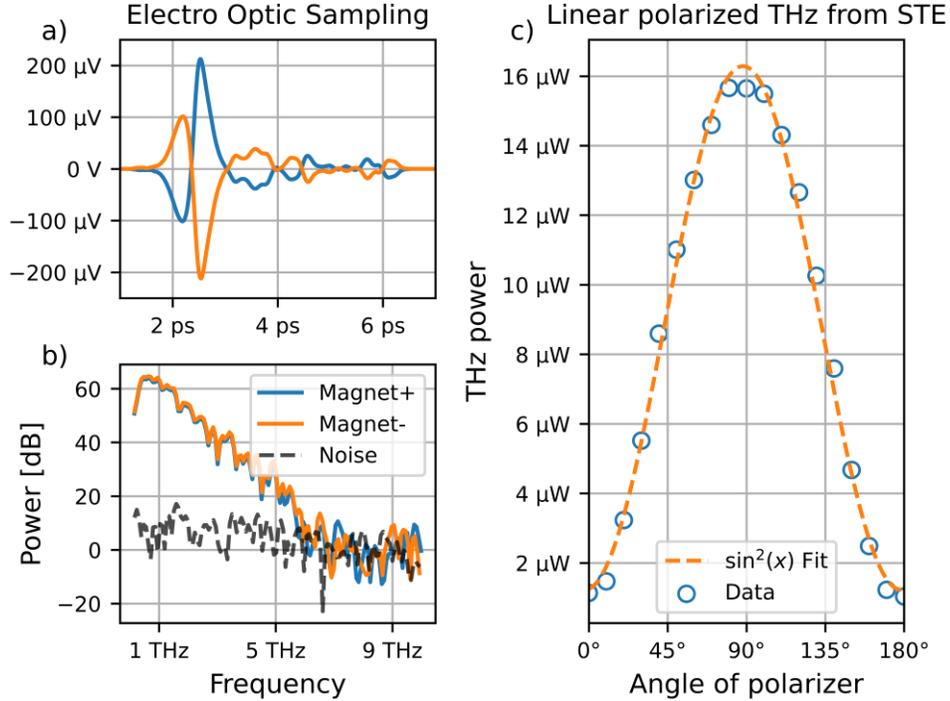

Fig. 6 a) Electro-optic sampling time trace for two different magnet configurations (Magnet+/Magnet-). Flipping the magnet switches the polarity of the THz trace in time domain. b) Corresponding power spectra are an accurate match to each other. c) A free space wire grid polarizer confirms the linear polarization of the emitted THz radiation, which is perpendicular to the magnetic field.

To further characterize the emitted THz radiation, a free-space wire-grid polarizer (P01, InfraSpecs) is placed in the collimated beam between the OAPs. Due to its limited clear aperture of 40 mm compared to the diameter of the OAPs with 75 mm, the THz power values in Fig. 6c) are not as high as in the previous subsections. Nevertheless, our data confirm the Malus law for linear polarization.

The EOS in the time and frequency domains is shown in Fig. 7 as a function of average pump power. The pump spot is the optimal chosen diameter of 1.8 mm ($1/e^2$) and the detection crystal is again a 200 µm thick GaP crystal. All measurements are conducted at the highest repetition rate of 400 kHz. As we can see, the spectral features are independent of pump power, and only the noise floor decreases, resulting in a maximum dynamic range of 45dB.

From the recorded EOS data, the peak-to-peak value in the time domain is extracted. Fig. 7c) shows this value versus pump power, and Fig. 7d) shows an integration of the squared time signal S(t) versus pump power. In Fig. 7c) we can identify two regimes when increasing the pump power: a linear regime at lower powers, followed by slow saturation and sub-linear dependence on the pump power at higher pump powers. We note that most setups presented so far are operating in the first-mentioned regime, meaning that the measured field scales linearly with pump power [20,21,45]. Extracting maximum power however is obtained by pumping beyond this point, and our EOS measurements suggest that this does not distort the emitted pulses, at least within the measured bandwidth. In Fig. 7d) the squared integration shows a linear proportionality for higher pump powers, which is the same trend presented in Fig. 5a). Here we can see the same behavior, the first regime is now a quadratic increase which merges into a linear proportionality. The EOS measurement has a higher sensitivity and can detect the quadratic increase at low pump powers better than the measurement with the power meter from Fig. 5a). It reproduces the measured THz power curve, confirming that no strong spectral distortions in the spectral window covered by our EOS are present.

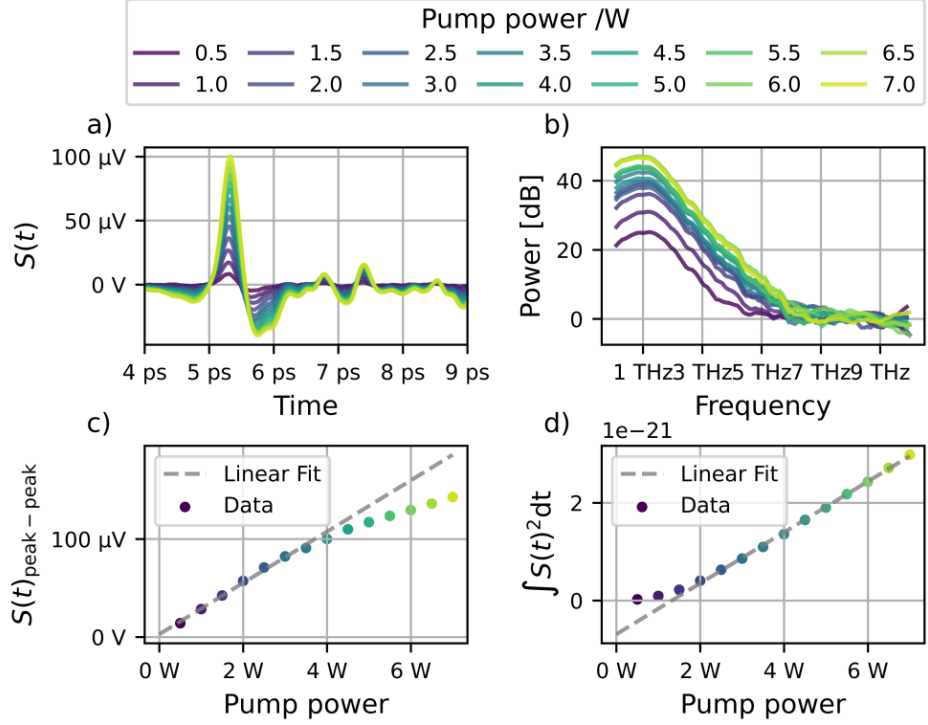

Fig. 7 a) Electro-optic sampling time trace for different pump powers (on the STE) with a pump diameter of 1.8 mm (1/e²). b) Corresponding power spectra, filtered, show a bandwidth of up to 6 THz at highest pump power. c) The peak-peak values of the THz time traces obtained by EOS are plotted versus pump power. The resulting voltage is proportional to the electric field for low pump powers. Data for linear fit up to 3.5 W. d) The time integral of the squared THz signal S(t) is a power value and proportional to the THz energy. For sufficient high pump powers, it scales only linear and not quadratically anymore. Data for linear fit from 2 W and above.

### 3.5 Discussion of STE output versus pump parameters

From the above experiments, the STE output power as a function of pump parameters can be summarized by the following scaling relation:

$$P_{\text{THz}} = C f_{\text{rep}} A_{\text{pump}} g\left(\frac{F_{\text{pump}}}{F_c}\right) \quad (1)$$

Here, C is a constant coefficient, $f_{rep}$ is the pump-pulse repetition rate, and $A_{pump}$ is the pump-spot area. The function $g(x)$ describes the scaling with regard to the pump fluence $x = F_{pump}/F_c$ relative to the critical fluence $F_c$. It captures three regimes. (i) In the weak-excitation regime (for small $x$), the emitted THz power grows quadratically with $x$. (ii) In the intermediate regime, the THz power grows linearly with $x$. (iii) At high fluences ($x \gtrsim 1$), the THz power saturates.

To qualitatively discuss the behavior of $g(x)$, we note that the THz electric field behind the STE scales like [46]

$$E \propto M_{\text{eq}}(T_0 + \Delta T_{\text{e,max}}) - M_{\text{eq}}(T_0) \qquad (2)$$

Here, $M_{\text{eq}}$ is the equilibrium magnetization of the STE, $T_0$ is the stationary STE temperature, and $\Delta T_{\text{e,max}}$ is the peak of the pump-induced increase of the electron temperature of the STE magnetic layer.

Therefore, two effects contribute to the fluence dependence of the STE: first, the temperature dependence of $M_{\text{eq}}$ [47] and, second, the temperature dependence of the electronic heat capacity of the STE stack [48]. The latter determines the relationship between the deposited pump fluence and the resulting $\Delta T_{\text{e,max}}$.

Within this simple model, we can rationalize the different observed regimes of $g(x)$ in Eq.(1): In regime (i), $E$ scales linearly with $\Delta T_{\text{e,max}}$, and $\Delta T_{\text{e,max}}$ is proportional to the pump fluence. In regime (ii), the electronic temperature enters regions in which the magnetic contribution to the electronic heat capacity becomes sizeable [48]. As a result, the change $\Delta T_{\text{e,max}}$ in electronic temperature scales sub-linearly with the absorbed pump fluence, whereas $E$ scales still approximately linearly with $\Delta T_{\text{e,max}}$. Finally, in regime (iii), the ferromagnetic material is driven into the state of total demagnetization [47], that is, the spin current source is fully exhausted by the action of the powerful pump pulse.

We emphasize that this explanation is only qualitative. For example, we note that Eq.(2) was derived only for the low-fluence regime [46]. Thus, there may be modifications at higher pump fluences. A full explanation needs to extend Eq.(2) to arbitrary pump fluences.

*3.6 Variation of repetition rate and thermal behavior at optimal fluence*

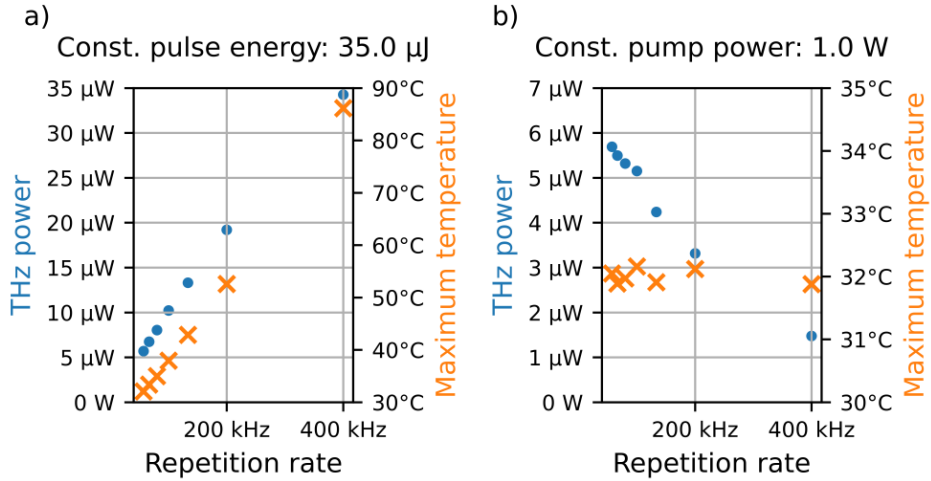

Fig. 8 a) Emitted THz power versus pump repetition rate for a constant pump energy of 35 µJ. The right axis, (in orange) displays the corresponding maximum surface temperature of the STE. The THz power and the maximum surface temperature show a linear increase with repetition rate. The temperature data is taken with thermal camera b) As panel a), but for a constant average power of 1 W. With increasing repetition rate, the pulse energy is lower and, thus, the THz generation efficiency decreases. The temperature data confirms the constant average power with a constant heat on the surface. The pump diameter is 1.8 mm ($1/e^2$).

To disentangle average power from pulse energy effects, we vary the repetition rate of the laser and use a constant pulse energy [Fig. 8a)] or a constant pump power [Fig. 8b)]. We focus here on the optimized case with 1.8 mm ($1/e^2$) laser spot size. Fig. 8a) shows the THz power increases linearly with the pump power on the STE at a constant pulse fluence. The time-averaged maximum surface temperature is recorded with the thermal camera. The temperature increase also scales linearly with the pump power and reaches a maximum of approximately 86°C at 7 W pump power (on the STE) and 400 kHz, resulting in $\Delta T \approx 65 K$ above room temperature. This indicates that the STE operates in a linear absorption regime. The cooling power at that operation point is 4.2 W, closely corresponding to an approximate 50% absorption and 35% transmission of the incident power as well as keeping the base temperature below room temperature. Together with the conclusions made in the previous paragraph, we can assume that this maximum temperature in the range of 86°C can be considered a good criterion for future work in other repetition rate regimes to stay in an efficient, non-saturated THz emission regime.

In Fig. 8b), the pump power is kept constant at 1 W for different repetition rates, which we achieve by varying the waveplate in the setup after each change of repetition rate to adjust the pump power to 1 W. The maximum THz power (and, thus, efficiency, since the pump power is kept constant) is at the highest pulse energy of 35 µJ. This observation confirms that the pump fluence is the main parameter relevant to power scaling of these emitters in our excitation regime. The temperature measurement on the right axis in Fig. 8b) confirms that the emitter operates in a linear absorption regime with constant pump power, as the temperature remains constant.

## 4. Conclusion and outlook

In conclusion, we explored THz generation in a tri-layer spintronic THz emitter on a sapphire substrate, used in reflection geometry and excited in the high-average-power and high-repetition-rate regime. As an excitation source, we use a commercial ultrafast fiber laser capable of delivering 7 W of average power impingent on the STE with variable repetition rate between 40 kHz and 400 kHz. The laser delivers 300 fs pulses, which we temporally compress externally to reach sub-30 fs for our study.

To the best of our knowledge, this is the first demonstration of an STE used in reflection geometry for improving average power handling. The corresponding improvement in heat handling enabled us to apply higher average powers than in previously demonstrated experiments, even in a regime of high repetition rate, where the pulse energies are low, and thus the spot sizes are correspondingly small to reach optimal conversion efficiency. Furthermore, optimal cooling allows to operate at overall lower temperatures which is favorable for the stability and long-term degradation of the emitter.

Varying the repetition rate of the source allowed us to disentangle the effects of average power and pulse energy to study thermal effects and compare the reflection geometry to the more standard transmission geometry. We show that even at the highest repetition rate available, the good heat extraction in the reflection case allows us to operate in a repetition-rate-independent regime, in which the ideal fluence for THz generation is not affected by the high average power. The obtained optimal fluence of 5 mJ/cm$^2$ also confirms very recent results in the low average power regime by Kumar et al. [41].

We believe these results are a first important milestone in average power scaling of metallic spintronic THz emitters. Ultrafast laser systems with multi-kW average powers are nowadays available [26–29] and the thermal exploration presented here allows us to predict how to make use of these advanced systems in the near future. In particular, our current result paves the way for similar results at MHz repetition rates operating at comparable pulse energies but much higher average powers, by optimizing the cooling geometry, for example using better substrates, or cooling at lower temperatures. This would enable the demonstration of scalable ultra-broadband THz sources with the potential of high dynamic range detection, which remains an ongoing challenge for THz-TDS sources.

**Acknowledgments**

We acknowledge support by the Open Access Publication Funds of the Ruhr-Universität Bochum. The results are funded by the Deutsche Forschungsgemeinschaft (DFG, German Research Foundation) under Germanys Excellence Strategy – EXC-2033 – Projektnummer 390677874 - RESOLV. This research is funded by the Deutsche Forschungsgemeinschaft (DFG, German Research Foundation) under Project-ID 287022738 TRR 196 for Project M01, and in part by the Alexander von Humboldt Stiftung (Sofja Kovalevskaja Preis). These results are part of a project that has received funding from the European Research Council (ERC) under the European Union's Horizon 2020 research and innovation programme (grant agreement No. 805202 - Project Teraqua). M. K. and G. J. acknowledge support by the German Research Foundation (SFB TRR 173 Spin+X #268565370, projects A01 and B02) and from the Horizon 2020 Framework Program of the European Commission under FET-Open grant agreement no. 863155 (s-Nebula). TSS and TK thank the German Research Foundation (DFG) for funding through SFB TRR 227 "Ultrafast spin dynamics" (projects A05 and B02) and the European Union for support through the ERC H2020 CoG project TERAMAG/Grant No. 681917.

**Disclosures**

TSS and TK are shareholders of TeraSpinTec GmbH and TSS is an employee of TeraSpinTec GmbH.

**Data Availability Statement**

Data underlying the results presented in this paper are not publicly available at this time but can be obtained from the authors upon reasonable request.